\documentclass[a4paper,fleqn,usenatbib]{mnras}

\usepackage{newtxtext,newtxmath}

\usepackage[T1]{fontenc}
\usepackage{ae,aecompl}

\usepackage{graphicx}	
\usepackage{amsmath}	
\usepackage{stmaryrd}
\usepackage{pdflscape}

\usepackage[OT2,OT1]{fontenc}
\newcommand\cyr
{
	\renewcommand\rmdefault{wncyr}
	\renewcommand\sfdefault{wncyss}
	\renewcommand\encodingdefault{OT2}
	\normalfont
	\selectfont
}
\DeclareTextFontCommand{\textcyr}{\cyr}

\title[\texttt{FastChem 2}]{\texttt{FastChem 2}: An improved computer program to determine the gas-phase chemical equilibrium composition for arbitrary element distributions}

\author[J. W. Stock et al.]{
Joachim W. Stock,\thanks{E-mail: joachimstock14@gmail.com}
Daniel Kitzmann,$^{1}$\thanks{E-mail: daniel.kitzmann@csh.unibe.ch}
A. Beate C. Patzer$^{2}$
\\
$^{1}$Center for Space and Habitability, University of Bern, Gesellschaftsstrasse. 6, 3012 Bern, Switzerland\\
$^{2}$Zentrum f\"ur Astronomie und Astrophysik (ZAA), Technische Universit\"at Berlin (TUB), Hardenbergstr. 36, 10623 Berlin, Germany
}

\date{Accepted 2022 September 12. Received 2022 September 12; in original form 2022 June 16}

\pubyear{2022}

\begin{document}
\label{firstpage}
\pagerange{\pageref{firstpage}--\pageref{lastpage}}
\maketitle

\begin{abstract}
The computation of complex neutral/ionised chemical equilibrium compositions is invaluable to obtain scientific insights of, for example, the atmospheres of extrasolar planets and cool stars.
We present \texttt{FastChem 2}, a new version of the established semi-analytical thermochemical equilibrium code \texttt{FastChem}.
Whereas the original version is limited to atmospheres containing a significant amount of hydrogen, \texttt{FastChem 2} is also applicable to chemical mixtures dominated by any other species, such as CO$_2$ or N$_2$. 
The new C++ code and an optional Python module are publicly available under the GPLv3 license at \url{https://github.com/exoclime/FastChem}.
The program is backward compatible so that the previous version can be easily substituted.
We updated the thermochemical database by adding HNC, FeH, TiH, Ca$^-$, and some organic molecules.
In total 523 species are now in the thermochemical database including 28 chemical elements. The user can reduce the total number of species to, for example, increase the computation performance or can add further species if the thermochemical data are available.
The program is validated against its previous version and extensively tested over an extended pressure-temperature grid with pressures ranging from $10^{-13}\,\mathrm{bar}$ up to $10^3\,\mathrm{bar}$ and temperatures between $100\,\mathrm{K}$ and $6000\,\mathrm{K}$.
\texttt{FastChem 2} is successfully applied to a number of different scenarios including nitrogen, carbon, and oxygen-dominated atmospheres as well as test cases without hydrogen and helium.
Averaged over the extended pressure-temperature grid \texttt{FastChem 2} is up to 50 times faster than the previous version and is also applicable to situations not treatable with version 1.
\end{abstract}



\section{Introduction}
To gain a detailed scientific understanding of astrophysical objects such as, e.g., the atmospheres of extrasolar planets, protoplanetary disks, or cool stars, the computation of their often complex chemical compositions is indispensable.
Therefore, rapid and efficient computer programs are needed.
The chemical equilibrium situation is thermodynamically determined by the minimum of the Gibbs free energy of the system \citep[see e.g.][]{Den55,Ari69}.
In combination with the conservation of elements the resulting chemical equilibrium (CE) composition can be numerically calculated by minimising the Gibbs free energy directly \citep[e.g.][]{Whi57,Whi58} or by employing the law of mass action\footnote{The empirical law of mass action \citep{Gul64,Gul67,Gul79} can be mathematically derived by minimising the Gibbs free energy \citep[see e.g.][]{Ari69}.} \citep[e.g.][]{Bri47}.
Detailed descriptions of various numerical methods to compute CE compositions can be found in the textbooks by \citet{Zeg70} or \citet{Smi82}, for example.

Assuming CE, \citet{Rus34} was able to estimate the chemical composition of stellar atmospheres with a then laborious procedure restricted to diatomic molecules and a strictly hierarchical structure of the chemical element abundances.
Russell's method became more feasible with the use of electronic computers and has been further developed to account for larger molecules and more species by utilising the Newton-Raphson method.
Such methods have been applied to investigate, for example, cool stellar atmospheres \citep{Var66,Tsu73,Joh82}, circumstellar dust shells of asymptotic giant branch (AGB) stars \citep{Gai84,Gai86,Gai87,Dom90,Win94,Fer01}, brown dwarfs \citep{Bur02,Mar02,Hel08mnras,Hel08apj}, and to some extend the atmospheres of extrasolar planets \citep{Mad11,Kat14,Mor15}. 
To further explore the chemical composition of planetary atmospheres and atmospheres of brown dwarfs under CE conditions, codes such as CONDOR \citep{Lod93}, TECA \citep{Ven12}, TEA \citep{Ble16}, GGChem \citep{Woi18}, and \texttt{FastChem} \citep{Sto18} were developed.

\texttt{FastChem} is so far applied to situations, where the CE approximation appears to be reasonable, that is, the chemical time-scale is considerably shorter than the dynamical time-scale and non-CE processes such as, e.g., photochemical processes or cosmic ray induced processes can be neglected. The applications of \texttt{FastChem} include for example particular atmospheric regions of the extrasolar planets 
KELT-9b \citep{Hoe18,Hoo18,Kit18,Hoe19,Fis20,Pin20,Won20}, KELT-20b/MASCARA-2b \citep{Hoe20a,Nug20mnras,Rai21,Joh22}, 
WASP-19b \citep{Sed21mnras},
WASP-33b \citep{Nug20apjl,Nug21},
WASP-76b \citep{Sei21,Sav22},
WASP-121b \citep{Hoe20b},
WASP-189b \citep{Pri22}, 
TOI-1518b \citep{Cab21}, 
TOI-2109b \citep{Won21},
HD149026b \citep{Ish21},
HD332231b \citep{Sed22},
and HAT-P-70b \citep{Bel22}.
In particular, \texttt{FastChem} is used to determine the chemical composition which is essential for self-consistent modelling of their respective atmospheres.
As a result, these models helped for example to detect or confirm neutral and ionised species such as, e.g., Ca, Ca$^+$, Cr, Cr$^+$, Fe, Fe$^+$, H, OH, Mg, Mn, Na, Ni, TiO, Sc$^+$, Ti, Ti$^+$, V, and Y$^+$ in the atmospheres of extrasolar planets. 
\texttt{FastChem} is also employed for comparison with non-equilibrium conditions, e.g., to study the impact of photochemistry, transport, emissions \citep{Men18,Cha21,Itc22}, and for validation of photochemical models \citep{Tsa18,Hob21}.
Moreover, \texttt{FastChem} has been combined with the exoplanet atmosphere model \texttt{HELIOS} \citep{Mal17,Mal19aj}, the opacity calculator \texttt{HELIOS-K} \citep{Gri15}, and/or the observation simulator \texttt{Helios-o} \citep{Bow19} in various applications \citep[e.g.][]{Lin19,Mal19apj,Gre20,Mor20,Ore20,Tab20,Fos21,Zil21,Dei22,Fei22,Guz22,Her22,Mar22,Zie22}.
Other applications of \texttt{FastChem} are the combination with the spectral characterisation tool \texttt{petitRADTRANS} \citep{Mol19} and as a plug-in for the retrieval code TauREx 3.1 \citep{Alr21,Alr22}.
The recently developed gas opacity generator \textsc{Optab} \citep{Hir22} is also compatible with \texttt{FastChem}.
Moreover, \texttt{FastChem} was coupled to the photochemical kinetics code \texttt{VULCAN} \citep{Tsa17,Shu20,Zil20,Das22} to provide initial values.

Here, we present a new version of \texttt{FastChem} henceforth called \texttt{FastChem 2}.
In contrast to the previous version (\texttt{FastChem 1}), \texttt{FastChem 2} does not rely on the presence of hydrogen as a reference element.
The basic equations are now written in a way that neither a specific reference element nor a conversion between the gas pressure and the total hydrogen nuclei number density are explicitly required for the calculation.  
Therefore, the new program can be applied to chemical element compositions even in total absence of hydrogen.
As another improvement, we now follow a semi-analytical approach for the calculation of the electron number density, if the absolute value of the ionisation degree of a species is less or equal one.
However, if the user provides thermochemical data including species with higher ionisation degrees, \texttt{FastChem 2} switches automatically to the method used in the previous version.
All these changes lead to a significant increase in computational performance, enabling a wider range of applications with higher speed.

In addition, some new species are added and thermochemical data are updated in this study.
These include hydrides, phosphides, sulphides, and some organic molecules with potential relevance to atmospheric chemistry. 
Finally, we provide an optional Python package \texttt{pyfastchem} that allows \texttt{FastChem 2} to be called directly from within Python scripts.
\texttt{FastChem 2} is backward compatible and the previous version can be easily replaced.

\section{Method}
In accordance with \citet{Sto18} we denominate the set of all chemical elements with $\mathcal{E}=\left\lbrace E_1,\ldots,E_{\left|\mathcal{E}\right|}\right\rbrace$ and the set of all species in the gas phase with $\mathcal{S}=\left\lbrace S_1,\ldots, S_{\left|\mathcal{S}\right|}\right\rbrace$. $\mathcal{S}$ includes for example free atoms, molecules, and ions whereas $\mathcal{E}$ includes solely chemical elements.
$\left|\mathcal{E}\right|$ is the total number of chemical elements and $\left|\mathcal{S}\right|$ is the total number of species treated. 
Because $\mathcal{E}\subset\mathcal{S}$, $\left|\mathcal{S}\right|$ is larger than $\left|\mathcal{E}\right|$.
The subset $\mathcal{S}$ without the chemical elements in atomic form is denoted by $\mathcal{S}\setminus\mathcal{E}$.
The index 0 refers here to the electron, i.e., $\mathcal{E}_0$ is the set of all chemical elements and $\mathcal{S}_0$ the set of all species both including the electron. 
To determine the CE composition, dissociative equilibrium is assumed, where the dissociation reaction
\begin{equation}
S_i\leftrightharpoons\nu_{i0}E_0+\nu_{i1}E_1+\ldots+\nu_{ij}E_j+\ldots=\sum_{j\in\mathcal{E}_0}\nu_{ij}E_j
\label{eq:reac}
\end{equation}
is considered for each species $S_i\in\mathcal{S}\setminus\mathcal{E}$.
If $j\neq0$, the coefficients $\nu_{ij}$ of the stoichiometric matrix are nonnegative integer numbers.
For cations $\nu_{i0}$ is negative and for anions the respective coefficient $\nu_{i0}$ is positive.
Consider for example the cation X$^+$, the associated dissociation reaction is
\begin{equation}
\mathrm{X}^+\leftrightharpoons\mathrm{X} - \mathrm{e^-}
\end{equation}
and $\nu_{\mathrm{X^+}0}=-1$.
Likewise for the anion X$^-$, the dissociation reaction is given by
\begin{equation}
\mathrm{X}^-\leftrightharpoons\mathrm{X} + \mathrm{e^-}
\end{equation}
and $\nu_{\mathrm{X^-}0}=1$.

The sum of all nuclei of the reference element $\mathrm{r}$ per unit volume is by definition
\begin{equation}
n_\mathrm{\left\langle r\right\rangle }:=n_\mathrm{r}+\sum_{i\in \mathcal{S}\setminus\mathcal{E}}\nu_{i\mathrm{r}}n_i\ ,
\label{eq:elemconsref}
\end{equation}
where $n_\mathrm{r}$ is the number density of the free atoms of the reference element $\mathrm{r}$.
Note, that in contrast to \citet{Sto18}, the reference element is not necessarily set to be hydrogen.

The CE composition $\left\lbrace n_0,\ldots,n_{\left|\mathcal{S}_0\right|} \right\rbrace$ can be calculated using the law of mass action \citep{Gul64,Gul67,Gul79}\footnote{For a standard textbook approach see e.g. \citet{Den55}. \citet{Lun65} provides a more in-depth historical context.}
\begin{equation}
n_i=K_i\prod_{j\in\mathcal{E}_0}n_j^{\nu_{ij}}\ ,\qquad\forall i\in\mathcal{S}\setminus\mathcal{E}\ ,
\label{eq:loma}
\end{equation}
where $n_i$ is the number density of species $i$ and $K_i$ the temperature dependent mass action constant, in combination with the element and charge conservation equations
\begin{equation}
\epsilon_{\mathrm{r},j} n_\mathrm{\left\langle r\right\rangle }=n_j+\sum_{i\in \mathcal{S}\setminus\mathcal{E}}\nu_{ij}n_i\ ,\qquad\forall j\in\mathcal{E}_0\ ,
\label{eq:elemcons}
\end{equation}
where $\epsilon_{\mathrm{r},j}$ is the abundance of element $j\in\mathcal{E}_0$ relative to $\mathrm{r}\in\mathcal{E}$.
Assuming charge neutrality, $\epsilon_{\mathrm{r},0}$ is zero in case of the electron.
Additionally, the number densities are constrained by the nonnegativity condition
\begin{equation}
n_i\geq 0\ ,\qquad i\in\mathcal{S}_0\ .
\end{equation}
So far condensates are not included in \texttt{FastChem}.
However, if the set of stable condensates obeying Gibbs' phase rule is known a priori, their impact on the gas phase can be included by adapting the element abundances $\epsilon_{\mathrm{r},j}$ appropriately.

\subsection{Input and output data}
The required input format is essentially the same as for \texttt{FastChem 1} \citep{Sto18} so user created input data files for \texttt{FastChem 1} can still be used for \texttt{FastChem 2}.
Note that we updated the thermochemical data and added some species (see section~\ref{sec:thermodata} for details). 
The temperature dependent mass action constants for each molecule or ion $S_i$ are fitted as before
\begin{equation}
\ln \bar{K}_i(T)=\frac{a_0}{T}+a_1 \ln T + b_0 + b_1 T + b_2 T^2\ ,\qquad\forall i\in\mathcal{S}\setminus\mathcal{E}\ ,
\label{eq:fit}
\end{equation}
with the five fit coefficients $a_0$, $a_1$, $b_0$, $b_1$, and $b_2$.
The natural logarithm of the dimensionless mass action constants are related to the change in the  Gibbs free energy of the dissociation reaction~(\ref{eq:reac}) with respect to the standard-state pressure $p^\minuso=1\,\mathrm{bar}$ ($=10^5\,\mathrm{Pa}$) via
\begin{equation}
\ln \bar{K}_i=-\frac{\Delta_\mathrm{r} G_i^\minuso(T)}{R\,T}\ .
\label{eq:deflnK}
\end{equation}
Although \texttt{FastChem 2} comes with its own thermochemical database, additional species can be easily included by the user if their mass action constants are available. Species or even elements can be removed by changing the respective input files, e.g., in order to further increase computational speed for specific purposes.

In contrast to \citet{Sto18}, the relative element abundances $\epsilon_{\mathrm{r},j}$ are automatically normalised to their total sum by \texttt{FastChem 2} via
\begin{equation}
\hat{\epsilon}_j=\frac{\epsilon_{\mathrm{r},j}}{\sum_{k\in\mathcal{E}}\epsilon_{\mathrm{r},k}}=\frac{10^{x_{\mathrm{r},j}}}{\sum_{k\in\mathcal{E}}10^{x_{\mathrm{r},k}}}\ ,\qquad j\in\mathcal{E}_0\ , 
\label{eq:defhateps}
\end{equation}
where the $x_{\mathrm{r},j}$ are provided by the user.
In stellar atmospheric theory $x_{\mathrm{r},\mathrm{H}}$ is usually set to $12$, where $\mathrm{r}=\mathrm{H}$ is the reference element by choice\footnote{If hydrogen is the chosen reference element, we use the notation $x_j$ instead of $x_{\mathrm{H},j}$.}.
Using \texttt{FastChem 2}, this convention 
does not necessarily need to be taken under consideration while creating an input data file as long as the relative element abundances are consistent.
It would not be of advantage to provide $\hat{\epsilon}_j$ directly, since this quantity refers only to a specific element mixture.

The output data file format of the stand-alone \texttt{C++} version is similar to the one generated by the previous version \texttt{FastChem 1}, that is a formatted file listing the gas pressure $p_\mathrm{g}$, the temperature $T$, the total gas density $n_\mathrm{g}$, the total number density of all nuclei in the gas phase $n_\mathrm{\left\langle g\right\rangle}$, and the number densities $n_i$ of all species $S_i\in\mathcal{S}_0$.
Furthermore, a monitor file is generated which includes information about convergence behaviour.
It is highly recommended to check this file after the calculation is completed.

\subsection{The \texttt{FastChem 2} algorithm}
\label{ssec:fastchem}
In most practical applications the total gas pressure $p_\mathrm{g}$ is given instead of $n_{\left\langle\mathrm{r}\right\rangle}$ or, assuming an ideal gas, the associated pressure $p_{\left\langle\mathrm{r}\right\rangle}=n_{\left\langle \mathrm{r}\right\rangle} k_\mathrm{B}T$, where $k_\mathrm{B}$ denotes the Boltzmann constant.
Therefore, $p_\mathrm{g}$ needs in general to be converted into $n_{\left\langle\mathrm{r}\right\rangle}$.
For chemical systems such as in H-He-dominated stellar atmospheres, $p_\mathrm{\left\langle H\right\rangle}$ can be reasonably well approximated by relatively simple analytic expressions or numerical schemes \citep[for examples see e.g.][]{Tsu73,Sha90,Gai14}.
In situations where higher precision is required or the number densities of the leading elements are not well known a priori, an iteration procedure is beneficial \citep{Sto18,Woi18}.

Here, we rephrase the governing equations to avoid the iteration in order to increase computational speed via elimination of $n_{\left\langle\mathrm{r}\right\rangle }$.
Furthermore, the new formulation permits to perform all calculations without the explicit use of a reference element $\mathrm{r}\in\mathcal{E}$ and hence the $p_\mathrm{g}$-$n_{\left\langle\mathrm{r}\right\rangle }$-conversion becomes obsolete.

\subsubsection{Preconditioning}
\label{sssec:preconditioning}
The total gas number density is given by
\begin{equation}
n_\mathrm{g}=\sum_{j\in\mathcal{E}_0}n_j+\sum_{i\in\mathcal{S}\setminus\mathcal{E}}n_i\ .
\label{eq:ntot}
\end{equation}
Equations (\ref{eq:elemcons}), (\ref{eq:elemconsref}), and (\ref{eq:ntot}) can be used to eliminate $n_\mathrm{\left\langle r\right\rangle }$ and the number density of the reference element $n_\mathrm{r}$ , yielding
\begin{equation}
\sum_{k\in\mathcal{E}_0}a_{jk} n_k=\epsilon_{\mathrm{r},j} n_\mathrm{g}-\sum_{i\in \mathcal{S}\setminus\mathcal{E}}\left[\nu_{ij}+\epsilon_{\mathrm{r},j}\left(1-\nu_{i\mathrm{r}} \right)\right]n_i\ ,
\label{eq:master1}
\end{equation}
where
\begin{equation}
a_{jk}=\delta_{jk}+\epsilon_{\mathrm{r},j}\left(1-\delta_{\mathrm{r}k} \right)
\label{eq:defajk}
\end{equation} 
and $\delta_{jk}$ denotes the Kronecker delta.
The left hand side of equation~(\ref{eq:master1}) is linear in the number densities of the elements $n_k$ and can be solved for $n_j$ ($j\in\mathcal{E}_0$).
\begin{equation}
n_j=\sum_{k\in\mathcal{E}_0}\bar{a}_{jk}\epsilon_{\mathrm{r},k} n_\mathrm{g}-\sum_{k\in\mathcal{E}_0}\sum_{i\in\mathcal{S}\setminus\mathcal{E}}\bar{a}_{jk}\left[\nu_{ij}+\epsilon_{\mathrm{r},j}\left(1-\nu_{i\mathrm{r}} \right)\right]n_i\ ,
\label{eq:master2}
\end{equation}
where
\begin{equation}
\bar{a}_{jk}=\delta_{jk}-\hat{\epsilon}_j\left(1-\delta_{\mathrm{r}k} \right)
\end{equation}
are the components of the inverse of the matrix $\left( a_{jk}\right)$.
Substituting $\bar{a}_{jk}$, equation~(\ref{eq:master2}) can be further simplified to
\begin{equation}
n_j=\hat{\epsilon}_j n_\mathrm{g}-\sum_{i\in \mathcal{S}\setminus\mathcal{E}}\left[\nu_{ij}+\hat{\epsilon}_j\sigma_i\right]n_i\ .
\label{eq:master2simple}
\end{equation}
with
\begin{equation}
\sigma_i=1-\sum_{j\in\mathcal{E}_0}\nu_{ij}\ .
\end{equation}
After utilising the law of mass action (equation~\ref{eq:loma}), we decompose the equation system~(\ref{eq:master2simple}) into a set of equations each in one variable, namely, $n_j$ ($j\in\mathcal{E}$) 
\begin{equation}
\hat{\epsilon}_j n_\mathrm{g}=n_j+\sum_{k=1}^{N_j} n_j^k\underset{\hat{\epsilon}_i=\hat{\epsilon}_j}{\underset{\nu_{ij}=k}{\sum_{i\in \mathcal{S}\setminus\mathcal{E}}}}\left[\nu_{ij} +\hat{\epsilon}_j\sigma_i\right]K_i\underset{l\neq j}{\prod_{l\in\mathcal{E}_0}}n_l^{\nu_{il}}+\bar{n}_j+n_{j,\mathrm{min}}
\label{eq:master3}
\end{equation}
similar to \citet{Sto18}, where
\begin{equation}
n_{j,\mathrm{min}}=\underset{\hat{\epsilon}_i<\hat{\epsilon}_j}{\sum_{i\in\mathcal{S}\setminus\mathcal{E}}}\left[\nu_{ij}+\hat{\epsilon}_j\sigma_i\right]n_i 
\end{equation}
accounts for the contribution of species which consist of elements of which at least one is less abundant than element $j$.
Moreover,
\begin{equation}
\bar{n}_j=\underset{\hat{\epsilon}_i=\hat{\epsilon}_j}{\sum_{i\in \mathcal{S}\setminus\mathcal{E}}}\hat{\epsilon}_j\sigma_i n_i
\label{eq:defnjbar}
\end{equation}
describes the contribution of species containing elements more abundant than element $j$.
Furthermore, we defined
\begin{equation}
\hat{\epsilon}_i=\min_{j\in\mathcal{E}}\left\lbrace \left. \hat{\epsilon}_j\right|\nu_{ij}\neq 0\right\rbrace\ ,\qquad i\in\mathcal{S}\setminus\mathcal{E} 
\end{equation}
and
\begin{equation}
N_j=\max_{i\in\mathcal{S}\setminus\mathcal{E}}\left\lbrace\left. \nu_{ij}\right| \hat{\epsilon}_i=\hat{\epsilon}_j\right\rbrace\ ,\qquad j\in\mathcal{E}\ . 
\end{equation}
Introducing the following coefficients
\begin{equation}
A_{j0}=\bar{n}_j+n_{j,\mathrm{min}}-\hat{\epsilon}_j n_\mathrm{g}\ ,
\label{eq:defAj0}
\end{equation}
\begin{equation}
A_{j1}=1+\underset{\hat{\epsilon}_i=\hat{\epsilon}_j}{\underset{\nu_{ij}=k}{\sum_{i\in \mathcal{S}\setminus\mathcal{E}}}} \left[1+\hat{\epsilon}_j\sigma_i\right]K_i\underset{l\neq j}{\prod_{l\in\mathcal{E}_0}}n_l^{\nu_{il}}\ ,
\label{eq:defAj1}
\end{equation}
and
\begin{equation}
A_{jk}=\underset{\hat{\epsilon}_i=\hat{\epsilon}_j}{\underset{\nu_{ij}=k}{\sum_{i\in \mathcal{S}\setminus\mathcal{E}}}} \left[k+\hat{\epsilon}_j\sigma_i\right]K_i\underset{l\neq j}{\prod_{l\in\mathcal{E}_0}}n_l^{\nu_{il}}\ ,\qquad k\geq 2
\label{eq:defAjk}
\end{equation}
equation~(\ref{eq:master3}) reduces to a polynomial equation
\begin{equation}
P_j(n_j):=\sum_{k=0}^{N_j}A_{jk}n_j^k=0
\label{eq:defPj}
\end{equation}
which is solved analytically if its degree $N_j$ is smaller or equal two and via the ordinary Newton-Raphson method \citep[e.g.][]{Deu04} in one dimension otherwise.

It might happen that $A_{j0}$ becomes positive in the early stages of the iteration.
In that case, we solve the full equation
\begin{equation}
\hat{\epsilon}_j n_\mathrm{g}=n_j+\sum_{k=1}^{N_j} n_j^k\underset{\nu_{ij}=k}{\sum_{i\in\mathcal{S}\setminus\mathcal{E}}}\left[k +\hat{\epsilon}_j\sigma_i\right]K_i\underset{l\neq j}{\prod_{l\in\mathcal{E}_0}}n_l^{\nu_{il}}
\label{eq:masteralt}
\end{equation}
for element $j$.
Since $A_{jk}$ can become a negative number, it can not be easily inferred that the objective functions $P_j(n_j)$ are strictly convex for $n_j\geq 0$.
However, it can be shown that for certain conditions $P_j(n_j)$ is equivalent to the strictly convex objective function used in \texttt{FastChem 1} (see Appendix~\ref{sec:append}).
Thus, the convergence of the Newton-Raphson method is guaranteed, if the calculated $n_\mathrm{\left\langle g\right\rangle}$ is sufficiently close to the solution.

\subsubsection{Determination of the electron density}
\label{sssec:electron}
The electron density is calculated assuming charge neutrality, i.e., $\epsilon_{\mathrm{r},0}=0$.
We distinguish between two cases.
If $\left|\nu_{i0}\right|\leq1$ for all $i\in\mathcal{S}$, it follows from equation~(\ref{eq:elemcons})
\begin{equation}
0 = n_0\left(1+\underset{\nu_{i0}=1}{\sum_{i\in\mathcal{S}\setminus\mathcal{E}}}K_i\prod_{j\in\mathcal{E}}n_j^{\nu_{ij}}\right)-\frac{1}{n_0}\underset{\nu_{i0}=-1}{\sum_{i\in\mathcal{S}\setminus\mathcal{E}}}K_i\prod_{j\in\mathcal{E}} n_j^{\nu_{ij}}
\end{equation}
which is then solved analytically.
Otherwise, the electron density is calculated from the sum of the ion densities
\begin{equation}
n_0=-\sum_{i\in \mathcal{S}}\nu_{i0}n_i
\end{equation}
\citep[see e.g.][]{Gai14}.
If cancellation of leading digits causes numerical problems, we first try to solve equation~(\ref{eq:masteralt}) for $j=0$ by employing the ordinary Newton-Raphson-method.
Therefore, we set the initial value $n_0^{(0)}=Z/\left(Z+1\right)n_\mathrm{g}$, where $Z=\left| \min_{i\in\mathcal{S}}\nu_{i0}\right|$.
If the Newton-Raphson method still fails to converge, we utilise the method of Nelder and Mead \citep{Nel65,Lag97} as described by \citet{Sto18}.
To avoid numerical oscillations the resulting electron density $n_0^{(\mu)}$ is modified according to
\begin{equation}
n_0^{(\mu)}\leftarrow\sqrt{n_0^{(\mu)}n_0^{(\mu-1)}}\ ,
\end{equation}
where $\mu$ is the iteration step.

\subsubsection{Computational procedure}
Since \texttt{FastChem 2} has been developed on the basis of the previous version \texttt{1} \citep{Sto18}, the computational procedures are clearly similar despite the rephrasing of the governing equations.
Specifically, the element conservation equations, in combination with the law of mass action, are solved one by one in descending order, which is automatically determined beforehand, starting with the most abundant element.
An iteration procedure ensures a consistent mathematical solution.
Due to the new coupling term $\bar{n}_j$ in equation~(\ref{eq:master3}) each step, except for step 1, underwent some changes in comparison to the previous version of $\mathtt{FastChem}$.
\begin{description}
	\item[\textbf{Step 1}]\hfill\\Initial values for the electron density $n_0^{(0)}$ and for the correction terms $n_{j,\mathrm{min}}^{(0)}$ are set and the logarithmic mass action constants $\ln K_i$ are calculated for a given temperature $T$.
	\item[\textbf{Step 2}]\hfill\\The number densities for all atomic species $n_j\ (j\in\mathcal{E}$) are calculated via equation~(\ref{eq:master3}) (or equation~(\ref{eq:masteralt}) if necessary) in descending order, starting with the most abundant element.
	$n_i$ and $\bar{n}_j$ are updated at once during this step according to equations~(\ref{eq:loma}) and (\ref{eq:defnjbar}).
\item[\textbf{Step 4}]\hfill\\$n_{j,\mathrm{min}}$ is updated.
	\item[\textbf{Step 5}]\hfill\\The electron density $n_0$ is calculated (see Section~\ref{sssec:electron}).
\end{description}

\subsection{Implementation details}
Like \texttt{FastChem 1}, the core of version \texttt{2} is also written in \texttt{C++}. 
Major updates have been made to increase computational performance. 
\texttt{FastChem 2} can easily be incorporated into any other astrophysical or atmospherical models by using the provided object class. 
For example, this is done in the retrieval code \texttt{Helios-r2} \citep{Kit20} that can directly use \texttt{FastChem 2} during its forward model calculations. In addition to the actual \texttt{FastChem} code, the repository contains a \texttt{C++} stand-alone version that can be used for plain CE calculations and also showcases how \texttt{FastChem 2} can be coupled to other codes. 

As a major addition, we now provide a complete Python interface (\texttt{pyfastchem}) that allows the user to call \texttt{FastChem} directly from within a Python script. 
The \texttt{pyfastchem} package is also available on the Python Package Index (PyPI) which allows for a straightforward and easy installation of the Python module using the command: \texttt{pip install pyfastchem}. 
Besides the module itself, we also provide a series of scripts that show how \texttt{FastChem} is used within Python, also giving examples on how to, amongst others, change element abundances on the fly. 
The scripts can be easily adapted by the user for their special computational requirements.
By default the Python scripts produce the same general chemistry and monitor output files as the stand-alone \texttt{C++} version.
Furthermore, the Python version has also additional output capabilities, such as saving data of a specific subset of species or directly visualise the \texttt{FastChem} output.

The code is released under a GNU GPL 3.0 licence \citep[GPLv3, ][]{Gnu07} and is freely available on GitHub repository: \url{https://github.com/exoclime/FastChem}. The repository also includes a full user manual in pdf format that provides detailed information and instructions on how to compile the program and run the model. It also lists all available functions that allow the user to interact with \texttt{FastChem} and describes all required and optional input files.

\section{Additional and updated thermochemical data}
\label{sec:thermodata}
The thermochemical database which comes with $\mathtt{FastChem}$ has been refurbished.
In particular, we included the data of additional species especially of potential importance for stellar and planetary atmospheres from the literature and also updated the fit coefficients for the calculation of the mass action constants if newer and/or more precise data were at hand.
Table~\ref{tab:newmolecules} provides an overview of the changes in comparison to the previous version of the $\mathtt{FastChem}$ database (see also \citet{Sto18}, their Table~2).
\begin{table}
	\caption{Updated thermochemical data since the release of \texttt{FastChem 1} \citep{Sto18}. Newly added species are marked with an asterisk (*).} 
	\begin{tabular}{rll}
		\hline
		new & molecule       & reference \\
		\hline
		*&CH$_4$O$_2$        & \citet{Dor01}\\
		*&C$_2$H$_2$O$_2$    & \citet{Dor01}\\
		*&C$_2$H$_2$O$_4$    & \citet{Dor01}\\
		*&C$_2$H$_3$ClO$_2$  & \citet{Dor01}\\
		*&C$_2$H$_4$O$_3$    & \citet{Dor01}\\
		*&C$_2$H$_6$O$_2$    & \citet{Dor01}\\
		*&C$_2$NO            & \citet{Dor01}\\
		*&C$_3$N$_2$O        & \citet{Dor01}\\
		*&C$_4$H$_6$O$_4$    & \citet{Dor01}\\
		*&HNC                & \citet{Goo22}\\
		*&Ca$^-$             & \citet{Hoe19}\\
		 &CaH                & \citet{Bar16}\\
		 &ClH                & \citet{She04}\\
		 &CrH                & \citet{Bar16}\\
		 &CuH                & \citet{Bar16}\\
		 &HF                 & \citet{She04}\\
		*&FeH                & \citet{Bar16}\\
		 &MgH                & \citet{Bar16}\\
		 &MnH                & \citet{Bar16}\\
		 &NaH                & \citet{Bar16}\\
		 &NiH                & \citet{Bar16}\\
		 &HP                 & \citet{Lod99}\\
		 &HS                 & \citet{Lod04}\\
		*&TiH                & \citet{Bur05apj}\\
		 &HN                 & \citet{Goo22}\\
		 &HNO$_3$            & \citet{Dor03}\\
		 &H$_2$O$_2$         & \citet{Dor03}\\
		 &H$_2$SO$_4$        & \citet{Dor03}\\
		 &PH$_3$             & \citet{Lod99}\\
		 &PN                 & \citet{Lod99}\\
		 &NS                 & \citet{Lod04}\\
		 &SO$_2$             & \citet{Lod04}\\
		 &PS                 & \citet{Lod04}\\
		\hline
	\end{tabular}
	\label{tab:newmolecules}
\end{table}
The associated input file is backward compatible so it can also be applied by users of \texttt{FastChem 1}.

\paragraph*{Hydrides}
Since hydrogen is the most abundant chemical element in the universe, hydrides are expected to be a very important class of molecules in astrophysics.
\citet{Bar16} calculated partition functions and equilibrium constants for diatomic molecules based on improved data, specifically dissociation energies $D^\circ$ were taken from \citet{Hub79}, \citet{Cur91} and in particular \citet{Lou07}.
For CaH, CrH, CuH, MgH, MnH, NaH, and NiH we fitted \citet{Bar16}'s logarithmic mass action constants after conversion\footnote{Note, that \citet{Bar16} define the mass action constants using the Guldberg-Waage law in partial pressures, whereas in this work dimensionless constants $\bar{K}_i(T)$ are defined via the Gibbs free reaction energy (see equation~(\ref{eq:deflnK})).} according to equation~(\ref{eq:fit}).  
Furthermore we added FeH and TiH based on data from \citet{Bar16} and \citet{Bur05apj}, respectively, and the hydrogen halides HCl and HF based on data published by \citet{She04}.
Finally NH has been updated according to the data provided by \citet{Goo22}.

\paragraph*{Phosphides and Sulphides}
\citet{Lod99,Lod04} critically reevaluated the molecular data of the phosphides HP, PH$_3$, and NP and the sulphides HS, NS, SO$_2$, and PS to overcome inconsistencies in the JANAF thermochemical tables \citep{Cha98}\footnote{Note, that the data of most of these species so far included in the $\mathtt{FastChem}$ database already originate from other sources \citep[cf.][]{Sto18}.}.

\paragraph*{Organic molecules}
\citet{Dor01} provided molecular data for the organic species bromoacetic acid (CH$_2$Br--COOH), chloroacetic acid (CH$_2$Cl--COOH), oxopropanedinitrile (NC--CO--CN), glycolic acid (HO--CH$_2$--COOH), glyoxal (O=CH--CH=O), cyanooxomethyl (O\.CCN), oxalic acid (HO--CO--CO--OH), methyl hydroperoxide (CH$_3$--O--O--H), dimethyl peroxide (CH$_3$--O--O--CH$_3$) and diacetyl peroxide (CH$_3$--CO--O--O--CO--CH$_3$) not included so far.
These molecules can play a significant role in Earth's tropospheric (non-equilibrium) chemistry.
For example, methyl hydroperoxide is a possible end product of the methane oxidation process in the absence of NO and NO$_2$ \citep[see e.g.][]{Lev71,Mcc71,War78,Tho80,Log81,War88,Way00}.
Glyoxal is an observed ring fragmentation product of the reaction between toluene and hydroxyl \citep{Leb97,Way00}. 

Using the full \texttt{FastChem 2} database (28 elements, 523 species) in a gas mixture with solar photospheric element composition, glyoxal is the most abundant molecule of all the newly implemented organic species with the molecular formula C$_x$H$_y$O$_z$, where $x$, $y$, and $z$ are positive integers. Only formyl (H\.CO) and formaldehyde (CH$_2$O) are in general more abundant.

\paragraph*{Other molecules}
The molecular properties of nitric acid (HONO$_2$), sulphuric acid (H$_2$SO$_4$), and hydrogen peroxide (H$_2$O$_2$) are of particular interest for Earth's atmosphere.
Thus, the thermochemical data of these molecules were reevaluated by \citet{Dor03} using newer and improved data (e.g., taking into account the effect of internal rotation).
Apart from Earth,  H$_2$SO$_4$ was detected in the atmosphere of Venus \citep{Pol74,Sur86,Mar18,Tit18,Osc21}.
The presence of H$_2$O$_2$ was confirmed in Mars' atmosphere by \citet{Cla04} and \citet{Enc04}.

\paragraph*{Ions}
To investigate the chemical composition of the ultra-hot Jupiter KELT-9b, \citet{Hoe19} calculated fit coefficients compatible with $\mathtt{FastChem}$ for singly and doubly ionised atoms based on partition functions using data from \citet{Gur82}, \citet{Kur95}, \citet{Crc04}, \citet{Kra18}.
Furthermore, they added some anions from which we added Ca$^-$ to the database for the scenarios discussed in the following section.
Since doubly ionised cations are only relevant at extremely high temperatures and low pressures, they are not taken into account here.
The fit coefficients of the remaining ions are however listed by \citet{Hoe19} and can be easily included if the user is interested in them.

\section{Results and Discussion}
\subsection{Chemical composition}
\subsubsection{Planetary atmospheric element composition scenarios}
To demonstrate the functionality of \texttt{FastChem 2} for scenarios not dominated by hydrogen, we choose different C:H:N:O:Ar ratios as shown in Table~\ref{tab:abundances}.
The element abundance of argon $x_\mathrm{Ar}=6.40$ is the same for all scenarios.
For the remaining elements solar photospheric abundances \citep{Asp09} have been assumed, except for helium and neon, which have been removed from the present calculations, so that the total gas pressure $p_\mathrm{g}$ is not dominated by those noble gases.
\begin{table}
	\caption{Element distribution scenarios for different C, H, N and O abundance combinations. All other elements are set to solar abundances, except for He and Ne which have been omitted in these calculations. The C:H:N:O:Ar ratios of scenarios I, II, IVa and IVb roughly resemble the atmospheric C:H:N:O:Ar ratios of Earth, Titan, Mars and Mars with $x_\mathrm{C}$ and $x_\mathrm{O}$ swapped.} 
	\begin{tabular}{lrrrr}
		\hline
		scenario & $x_\mathrm{C}$ & $x_\mathrm{H}$ & $x_\mathrm{N}$& $x_\mathrm{O}$ \\
		\hline
		   I & 4.98 &      7.33 &  8.62 & 8.09\\
		  II & 9.25 &      9.86 & 11.05 & 6.54\\
		IIIa & 8.43 & $-\infty$ &  7.83 & 8.69\\
		IIIb & 8.69 & $-\infty$ &  7.83 & 8.43\\
		 IVa & 8.18 &      4.97 &  6.93 & 8.48\\
		 IVb & 8.48 &      4.97 &  6.93 & 8.18\\
		\hline
	\end{tabular}
	\label{tab:abundances}
\end{table}
The temperature has been varied between $500\,\mathrm{K}$ and $6000\,\mathrm{K}$ at constant total gas pressure $p_\mathrm{g}=5\times10^{-3}\,\mathrm{mbar}$.

In this paragraph four scenarios are discussed (I, II, IVa, IVb), two of them with $x_\mathrm{C}<x_\mathrm{O}$ (I, IVa), two of them with $x_\mathrm{C}>x_\mathrm{O}$ (II, IVb), two of them with $x_\mathrm{N}>\mathrm{max}\left\lbrace x_\mathrm{C},  x_\mathrm{O}\right\rbrace$ (I, II), and two of them fulfilling $x_\mathrm{N}<\mathrm{min}\left\lbrace x_\mathrm{C},  x_\mathrm{O}\right\rbrace$ (IVa, IVb).
For the calculation of the CE composition of the four scenarios, 26 elements and 519 species are taken into account.
The element distribution of the scenarios roughly correspond to the element distributions in the atmospheres of the solar system  planets Earth (I) and Mars (IVa), and Saturn's moon Titan (II).
Furthermore, a scenario IVb has been added, where the element distribution is the same as in scenario IVa, but with the values of $x_\mathrm{C}$ and $x_\mathrm{O}$ interchanged.
\begin{figure*}
	\resizebox{\hsize}{!}{\includegraphics{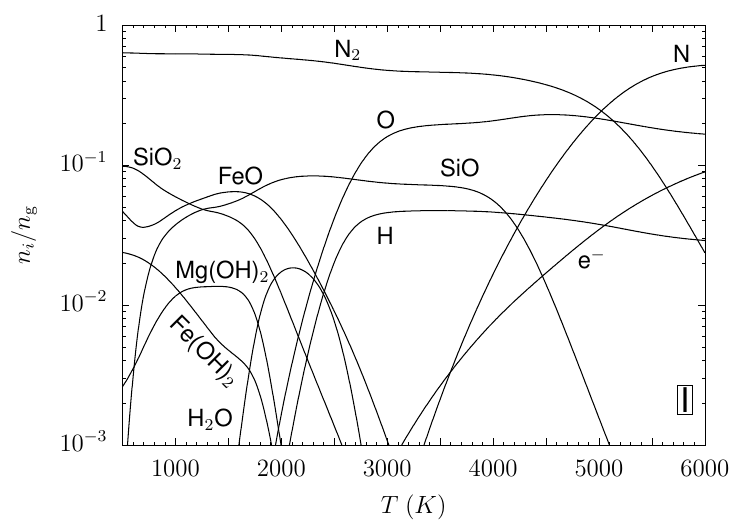}\includegraphics{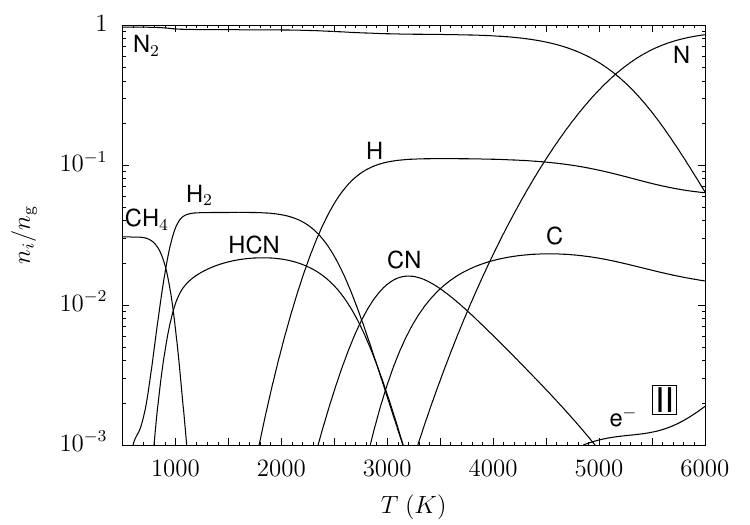}}\\	\resizebox{\hsize}{!}{\includegraphics{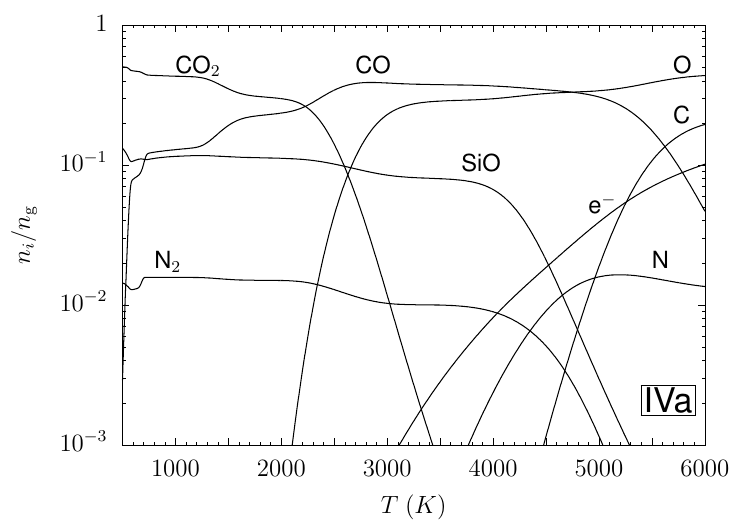}\includegraphics{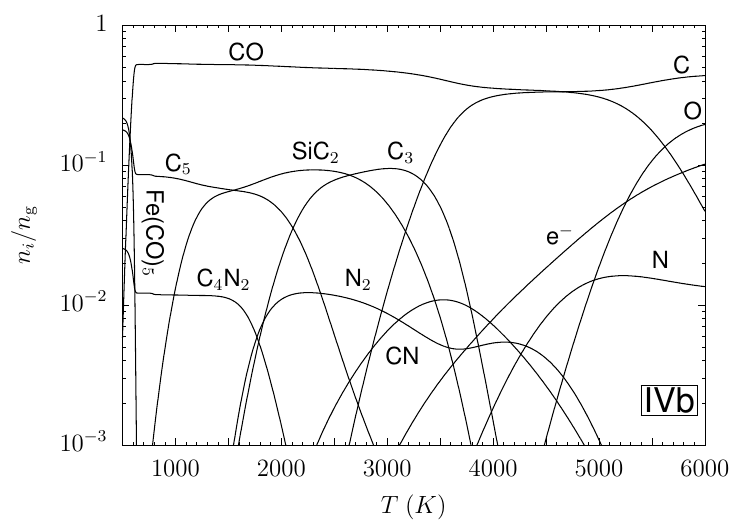}}
	\caption{Mixing ratios of the most abundant species, which include the elements C, H, N and/or O as function of temperature for a fixed total gas pressure $p_\mathrm{g}=5\,\mathrm{mbar}$ and different C:H:N:O:Ar ratios. The element abundances $x_\mathrm{C}$, $x_\mathrm{H}$, x$_\mathrm{N}$, and $x_\mathrm{O}$ for the scenarios I, II, IVa, and IVb are given in Table~1.}
	\label{fig:planets}
\end{figure*}

Figure~\ref{fig:planets} shows the most abundant species containing C, H, N, and/or O, as well as the electron abundance.
In the two left panels $x_\mathrm{C}$ is smaller than $x_\mathrm{O}$.
In the upper two panels $x_\mathrm{N}$ is larger than $\max\lbrace x_\mathrm{C},\ x_\mathrm{O}\rbrace$.
Unsurprisingly, the most prevalent species in scenarios I and II are N at higher temperatures ($T\gtrsim5000\,\mathrm{K}$) and N$_2$ at lower temperatures ($T\lesssim5000\,\mathrm{K}$).
In scenario I, $x_\mathrm{O}$ is about three orders of magnitude larger than $x_\mathrm{C}$, hence there is ample of oxygen available, forming numerous oxides such as FeO, SiO, SiO$_2$, and hydroxides, e.g., Fe(OH)$_2$ and Mg(OH)$_2$.
In scenario II, $x_\mathrm{C}$ exceeds $x_\mathrm{O}$ by about three orders of magnitude.
Due to the rich abundance of nitrogen and hydrogen, many hydrocarbons and carbon nitrides are present.
In scenarios IVa and IVb, $x_\mathrm{C}$ and $x_\mathrm{O}$ are in the same order of magnitude and nitrogen is less abundant than carbon and oxygen.
In the oxygen-rich case (IVa), carbon is locked in CO or CO$_2$ at lower temperatures. 
The remaining oxygen is predominantly bound in SiO.
In the carbon-rich case, more carbon is available, which leads to the presence of carbon compounds such as C$_3$, C$_5$, and SiC$_2$.
We note, these findings might be affected by including condensation processes or additional molecules.
At higher temperature, magnesium and iron are ionised.
In comparison to scenarios I, IVa, and IVb, the electron mixing ratio in scenario II is relatively low even at very high temperatures, due the low mixing ratio of magnesium and iron (see~Table~\ref{tab:abundances}).

\subsubsection{Two illustrating element compositions without hydrogen and helium}
We tested the code for two scenarios without the chemical elements hydrogen and helium, which by nature can not be successfully treated with \texttt{FastChem 1}. 
Therefore, we removed hydrogen and helium from the list of elements in the input file and used solar photospheric element abundances (scenario IIIa). Solar photospheric element abundances are used for scenario IIIb, whereby the numerical values of $x_\mathrm{C}$ and $x_\mathrm{O}$ are swapped.
Because of the removal of hydrogen and helium, only 402 species including the 26 elements remain in these two scenarios. 
\begin{figure}
	\resizebox{\hsize}{!}{\includegraphics{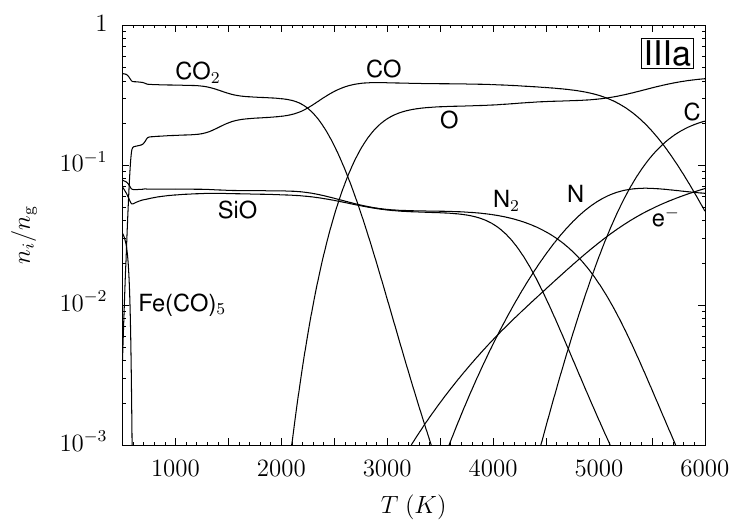}}\\
	\resizebox{\hsize}{!}{\includegraphics{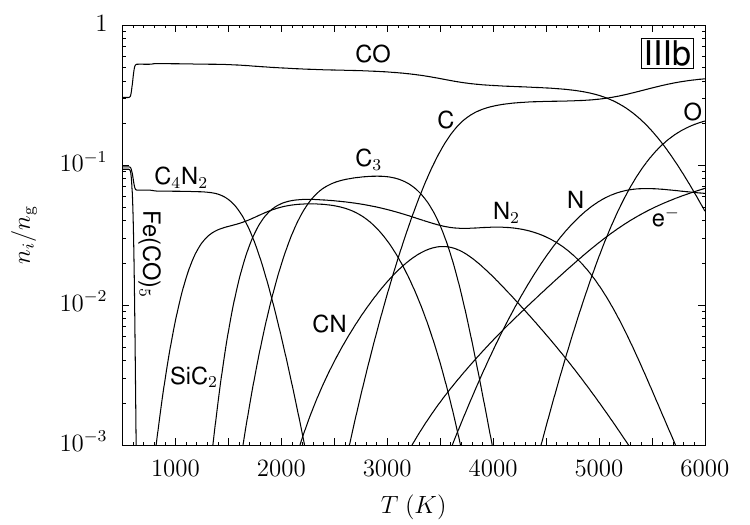}}\\	
	\caption{Volume mixing ratios of the most abundant carbons, nitrides, and oxides as function of temperature for a fixed total gas pressure $p_\mathrm{g}=5\,\mathrm{mbar}$. The element abundances $x_\mathrm{C}$, and $x_\mathrm{O}$ for the scenarios IIIa and IIIb are given in Table~1.}
	\label{fig:sun_wo_h_he}
\end{figure}
Figure~\ref{fig:sun_wo_h_he} shows the volume mixing ratios of the most abundant carbon, nitrogen, and oxygen compounds.
The upper panel shows the oxygen-rich scenario (IIIa), the lower panel shows the carbon-rich scenario (IIIb). 
Since hydrogen and helium have been excluded from these test calculations, carbon and oxygen are the most ample elements.
Hence CO, due to its large dissociation energy, has a special role in both scenarios, locking up the less abundant element (carbon or oxygen) over a large temperature range (see Figure~\ref{fig:sun_wo_h_he}).
In the oxygen-rich scenario (IIIa) the remaining oxygen can be found in species like O, SiO and CO$_2$, in the carbon-rich scenario (IIIb) the excess carbon is mostly bound in C$_3$, C$_4$N$_2$, SiC$_2$.
Evidently, there are no hydrocarbons or hydroxides present.
The electron densities are for both scenarios almost identical, since the element abundances for the main electron donating ions (Mg$^+$, Fe$^+$, Si$^+$) are the same.

\subsubsection{Chemical composition based on the evaporated DMM}
As a third example with neither hydrogen nor helium, we calculate the CE gas-phase composition of an evaporated \textbf{D}epleted \textbf{\underline{M}}id-\underline{O}cean \underline{R}idge \underline{B}asalt (MORB) \textbf{M}antle (DMM) using an atmospheric pressure-temperature structure.	
Therefore, the temperature is held constant at $T=2000\,\mathrm{K}$ and the total gas pressure $p_\mathrm{g}$ is varied between $10^{-1}\,\mathrm{bar}$ and $10^{-8}\,\mathrm{bar}$.
These conditions are comparable to those of a volatile-free atmosphere (see for example \citet{Sch09} and the related discussion of \citet{Mig11}).
Based on the chemical composition of the DMM's major elements \citep{Wor05}, we calculate the relative element abundances shown in Table~\ref{tab:abundancesmantle}.
\begin{table}
	\caption{Element abundances relative to oxygen derived from the chemical compostion of the DMM estimated by \citet{Wor05}}
	\begin{tabular}{llrllr}
		\hline
		Element & & $x_{\mathrm{O},j}$ & Element & & $x_{\mathrm{O},j}$ \\
		\hline
		O  & Oxygen    & 12.00 & Na & Sodium     & 9.37 \\
		Mg & Magnesium & 11.73 & Ni & Nickel     & 9.25 \\
		Si & Silicon   & 11.62 & Mn & Manganese  & 9.01 \\
		Fe & Iron      & 10.80 & Ti & Titanium   & 8.96 \\
		Ca & Calcium   & 10.50 & P  & Phosphorus & 8.17 \\
		Al & Aluminium & 10.34 & K  & Potassium  & 7.85 \\
		Cr & Chromium  &  9.62 &    &            &      \\
		\hline
	\end{tabular}
	\label{tab:abundancesmantle}
\end{table}
Note, that no hydrogen is present, so this scenario can not be treated with \texttt{FastChem 1}.
The total number of species composed of the 14 elements listed in Table~\ref{tab:abundancesmantle} is 80.
\begin{figure}
	\resizebox{\hsize}{!}{\includegraphics{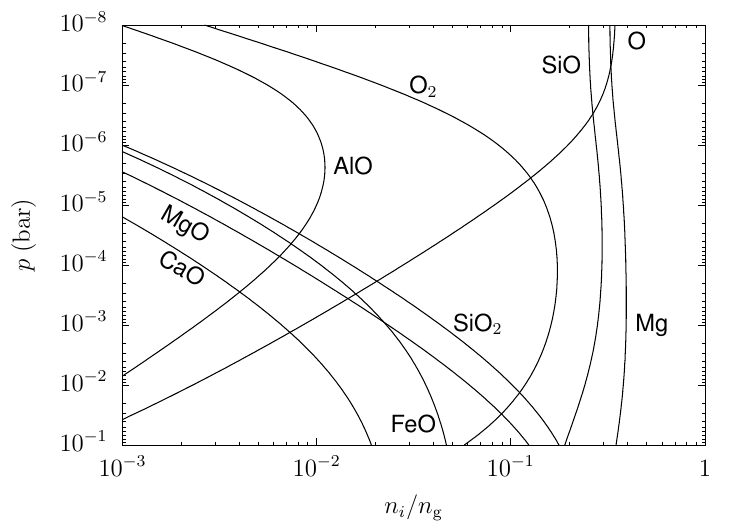}}	
	\caption{Volume mixing ratios of the most abundant oxides together with atomic magnesium as function of pressure for a constant temperature $T=2000\,\mathrm{K}$ using the chemical element composition given in Table~\ref{tab:abundancesmantle}.}
	\label{fig:evaporatingplanet}
\end{figure}
Figure~\ref{fig:evaporatingplanet} shows the mixing ratios of important species.
Since the DMM is mostly composed of MgO and SiO, the gas mixture associated with the DMM consists predominantly of Mg, SiO and at very low pressures of O. Note, that MgO has a very low dissociation energy in comparison to SiO. Other oxides such as SiO$_2$, MgO, FeO, CaO, and AlO as well as O$_2$ are also present, albeit in less amounts.

\subsection{Performance comparison with \texttt{FastChem 1}}
The performance differences between \texttt{FastChem} versions \texttt{1} and \texttt{2} are evaluated in several test calculations with both versions.
In particular, we calculate the chemical composition of $62\,500$ single pressure-temperature combinations on a $250 \times 250$ pressure-temperature grid considering all 28 elements and 523 species of the \texttt{FastChem 2} database. 
Pressures range from $10^{-13}\,\mathrm{bar}$ to $10^3\,\mathrm{bar}$ and temperatures from $100\,\mathrm{K}$ to $6000\,\mathrm{K}$ with $\Delta \log T$ and $\Delta \log p_\mathrm{g}$ held constant. 
Additionally, the hydrogen and helium element abundances $x_\mathrm{H}$ and $x_\mathrm{He}$ are varied in these tests. In this section, three distinct cases are discussed:
\begin{enumerate}
	\item In the first case only the hydrogen element abundance $x_\mathrm{H}$ is varied and the remaining element abundances $x_j$ are fixed at the solar photospheric value.
	\item In the second scenario the hydrogen and the helium element abundances $x_\mathrm{H}$ and $x_\mathrm{He}$ are varied while keeping the $x_\mathrm{H}$:$x_\mathrm{He}$-ratio constant.
	\item The last case is identical to the second but does not include ion chemistry.
\end{enumerate}
The computation time is obtained by taking the arithmetic mean of repeated calculations of the CE composition over the whole $p_\mathrm{g}$-$T$-grid.

\begin{figure}
	\resizebox{\hsize}{!}{\includegraphics{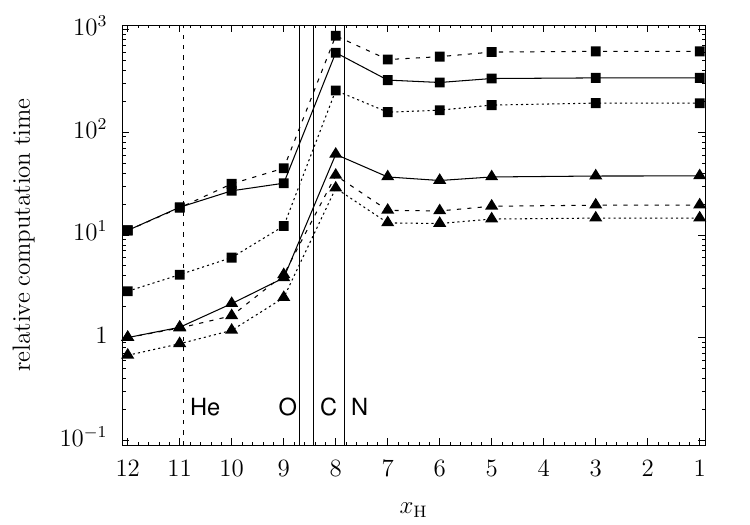}}
	\resizebox{\hsize}{!}{\includegraphics{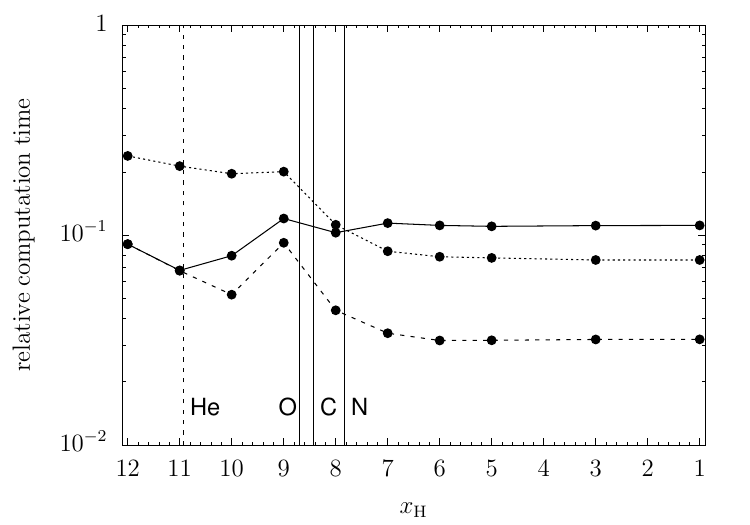}}	
	\caption{Upper panel: computation time relative to the reference scenario, using \texttt{FastChem 1} (squares) and \texttt{FastChem 2} (triangles), averaged over the $p_\mathrm{g}$-$T$-plane (see text). Lower panel: computation time of \texttt{FastChem 2} relative to \texttt{FastChem 1}. Solid lines refer to case (i), dashed lines to case (ii), and dotted lines to case (iii).}
	\label{fig:performance}
\end{figure}
Figure~\ref{fig:performance} shows the resulting computation time as function of the hydrogen element abundance $x_\mathrm{H}$. The upper panel depicts the average computation time for the complete $p_\mathrm{g}$-$T$-grid relative to the reference scenario (\texttt{FastChem 2} with ion chemistry for $x_\mathrm{H}=12$).
The lower panel shows the ratio between the runtimes of \texttt{FastChem 1} and \texttt{2}. For our test calculations, we use a desktop computer with an \textit{Intel i9-7960X} processor. The computation time\footnote{Note, that the computation time heavily depends on the computer hardware employed.} is about $15\,\mathrm{s}$ for the whole $p_\mathrm{g}$-$T$-grid. 

The results indicate that \texttt{FastChem 2} is substantially faster than \texttt{FastChem 1} for all cases (i) to (iii). 
Even at solar photospheric element abundances, the computation times of version \texttt{2} are only 10\% to 20\% of the other version.
For $x_\mathrm{H}>8$, the computation time of both versions increases with decreasing $x_\mathrm{H}$ for all cases.
This is mainly due to the decomposition of the system of equations~(\ref{eq:master2simple}) into a set of coupled non-linear equations~(\ref{eq:master3}) in one variable each. 
The resulting competition of other chemical elements such as carbon, nitrogen, and oxygen with hydrogen
causes the coefficient $A_{j0}$ to be positive and it becomes necessary to use equation~(\ref{eq:masteralt}) (see section~\ref{sssec:preconditioning}).
This effect reaches its maximum when $x_\mathrm{H}$ is of the same order of magnitude as $x_\mathrm{C}$, $x_\mathrm{N}$, and $x_\mathrm{O}$ resulting in a longest computation time at $x_\mathrm{H}\approx8$.
For $x_\mathrm{H}<8$, the computation time decreases slightly and then remains relatively constant for decreasing $x_\mathrm{H}$ with the chemistry now dominated by oxygen.

If $x_\mathrm{H}$ is larger than the solar photospheric helium abundance $x_\mathrm{He,solar}$, the computation time for cases (i) and (ii) is essentially at level.
If $x_\mathrm{H}$ is smaller than $x_\mathrm{He,solar}$, Figure~\ref{fig:performance} (upper panel) indicates qualitative differences.
The computation time of \texttt{FastChem 1} in case (ii) is larger than in case (i).
For low hydrogen abundances $n_\mathrm{g}$ is determined by $n_\mathrm{He}$ in version \texttt{1}.
Hence, in contrast to case (ii), the conversion between gas pressure $p_\mathrm{g}$ and hydrogen nuclei density $n_\mathrm{\left\langle H\right\rangle}$ is trivial for sufficiently low temperatures \citep[for details see][their section 2.3]{Sto18}.
	
An opposite behaviour can be seen, when applying \texttt{FastChem 2}. 
Here, the computation time of case (i) is larger than of case (ii).
The rationale behind this effect is different from \texttt{FastChem 1}, since here no $p_\mathrm{g}$-$n_\mathrm{\left\langle H\right\rangle}$-conversion is necessary.
By inspection of equation~(\ref{eq:master3}) it can be recognised that the main difference between case (i) and (ii) is the value of $\hat{\epsilon}_j$. 
For all chemical elements, except of helium, it follows from equation~(\ref{eq:defhateps}), that $\hat{\epsilon}_j$ is in case (ii) larger than in case (i).
Hence, the summands in equation~(\ref{eq:master3}), which include $\hat{\epsilon}_j$, such as the correction term $n_{j,\mathrm{min}}$, have a larger impact on the mathematical solution.
Consequently, more iteration steps are needed until convergence is achieved, resulting in an increase in computation time.
This is not the case for helium. 
However, the percentage in computation time of helium in comparison to the remaining elements is relatively modest.

The exclusion of the ion chemistry (case (iii)) leads to an increased performance between a factor $3.2$ and $5.2$ in version \texttt{1}.
In \texttt{FastChem 2} the performance gain is between $32\%$ and $67\%$.

With less than 3\% of the previous version's runtime at very low hydrogen and helium abundances, \texttt{FastChem 2} requires only a small fraction of the computational cost, needed by \texttt{FastChem 1}, to perform the same calculations.

\section{Summary}
We present a new updated version of \texttt{FastChem}, called \texttt{FastChem 2}, for the efficient and computationally fast calculation of chemical gas phase equilibria.
We modified the original \texttt{FastChem}-algorithm, so it can handle arbitrary element compositions.
Additionally, it can deal now with situations which are not dominated by hydrogen or helium.
We added some new species potentially relevant to atmospheric science and updated the thermochemical data used by \texttt{FastChem 2}.
The code is validated against \texttt{FastChem 1} and its functionality is demonstrated on several examples with different element composition over a wide range of pressure and temperature values.
A performance comparison shows that \texttt{FastChem 2} is significantly faster than \texttt{FastChem 1} up to a factor 50 in computation time.
The program is coded in object oriented \texttt{C++}, but it can be optionally called from within Python scripts using the new \texttt{pyfastchem} package.
\texttt{FastChem 2} and \texttt{pyfastchem} are open source and publicly available at github (\url{https://github.com/exoclime/FastChem}) under the GNU General Public License version 3 \citep{Gnu07}.

\section*{Acknowledgements}
DK acknowledges financial and administrative support by the Center for Space and Habitability and the PlanetS National Centre of Competence in Research (NCCR).

\section*{Data availability}
The data underlying this article are available in the FastChem GitHub repository, at \url{https://github.com/exoclime/FastChem}.




\bibliographystyle{mnras}
\bibliography{references} 



\appendix
\section{Treatment of numerical overflow}
In some cases numerical overflow could occur, mainly because of the logarithmic mass action constant (equation~(\ref{eq:deflnK})).  
This might happen for instance at very low temperatures $T$ or large values of Gibbs reaction energies $\Delta_\mathrm{r} G_i^\minuso(T)$.	
In order to avoid numerical overflow, equation~(\ref{eq:defPj}) can be optionally multiplied with a scaling factor $e^{-\psi_j}$
\begin{equation}
e^{-\psi_j}P_j(n_j)=\sum_{k=0}^{N_j}e^{-\psi_j}A_{jk}n_j^k=\sum_{k=0}^{N_j}\hat{A}_{jk}n_j^k=0\ ,
\end{equation}
with 
\begin{equation}
\psi_j:=\max_{i\in\mathcal{S}\setminus\mathcal{E}}\left(\ln K_i+\underset{l\neq j}{\sum_{l\in\mathcal{E}_0}}\nu_{il}\ln n_l \right) -\xi_j
\end{equation}
and $\xi_j\geq 0$ a non-negative constant which depends on the computer system and can be optionally specified by the user.
The default value is 0.
The modified coefficients are
\begin{equation}
\hat{A}_{j0}=e^{-\psi_j}A_{j0}=e^{-\psi_j}\left(\bar{n}_j+n_{j,\mathrm{min}}-\hat{\epsilon}_j n_\mathrm{g}\right)\ ,
\end{equation}
\begin{equation}
\begin{split}
\hat{A}_{j1}&=e^{-\psi_j}A_{j1}\\&=e^{-\psi_j}+\underset{\hat{\epsilon}_i=\hat{\epsilon}_j}{\underset{\nu_{ij}=k}{\sum_{i\in \mathcal{S}\setminus\mathcal{E}}}}\left[1+\hat{\epsilon}_j\sigma_i\right]\exp\left\lbrace \ln K_i+\underset{l\neq j}{\sum_{l\in\mathcal{E}_0}}\nu_{il} \ln n_l-\psi_j\right\rbrace\ ,
\end{split}
\end{equation}
and
\begin{equation}
\begin{split}
\hat{A}_{jk}&=e^{-\psi_j}A_{jk}\\&=\underset{\hat{\epsilon}_i=\hat{\epsilon}_j}{\underset{\nu_{ij}=k}{\sum_{i\in \mathcal{S}\setminus\mathcal{E}}}} \left[k+\hat{\epsilon}_j\sigma_i\right]\exp\left\lbrace \ln K_i+\underset{l\neq j}{\sum_{l\in\mathcal{E}_0}}\nu_{il}\ln n_l-\psi_j\right\rbrace
\end{split} 
\end{equation}
for all $k\geq 2$.

\section{Alternative derivation of equation~(16)}
In this section, we outline an alternative derivation of equation~(\ref{eq:master2simple}).
This derivation is slightly shorter, but probably less intuitive as the one presented in section~\ref{sssec:preconditioning}.
Instead of using a reference element $\mathrm{r}$, equation~(\ref{eq:master2simple}) can also be derived by making use of the total number density of atomic nuclei
\begin{equation}
n_{\left\langle\mathrm{g}\right\rangle}=\sum_{j\in\mathcal{E}}n_j+\sum_{j\in\mathcal{E}}\sum_{i\in \mathcal{S}\setminus\mathcal{E}}\nu_{ij}n_i\ .
\label{eq:nuclei}
\end{equation}
The element conservation can then be expressed via
\begin{equation}
\hat{\epsilon}_j n_{\left\langle\mathrm{g}\right\rangle}=n_j+\sum_{i\in \mathcal{S}\setminus\mathcal{E}}\nu_{ij}n_i\ ,\qquad j\in\mathcal{E}\ .
\label{eq:elemconsnuclei}
\end{equation}
By eliminating $\sum_{{j\in\mathcal{E}}}n_j$ and $n_{\left\langle\mathrm{g}\right\rangle}$ from equations~(\ref{eq:ntot}), (\ref{eq:nuclei}), and (\ref{eq:elemconsnuclei}), one easily obtains equation~(\ref{eq:master2simple}) after some rearrangement.
Note that $\hat{\epsilon}_j$ is defined by equation~(\ref{eq:defhateps}) as the normalised relative element abundance.
In context of equation~(\ref{eq:elemconsnuclei}), $\hat{\epsilon}_j$ can also be understood as the element abundance relative to the total number density of all atomic nuclei.

\section{On the objective functions utilised in \texttt{FastChem 1} and \texttt{FastChem 2}}
\label{sec:append}
Here, we point out the similarities between the objective functions used in \texttt{FastChem 1} and \texttt{FastChem 2}.
Starting from equation~(\ref{eq:master2simple}),
let
\begin{equation}
F_j(n_j):=n_j+\sum_{i\in\mathcal{S}\setminus\mathcal{E}}\left[ \nu_{ij}+\hat{\epsilon}_j\sigma_i\right]n_i-\hat{\epsilon}_j n_\mathrm{g}
\label{eq:objfunc}
\end{equation}
be an objective function.
The sum in equation~(\ref{eq:objfunc}) can be split yielding
\begin{equation}
F_j(n_j)=n_j
+\underset{\hat{\epsilon}_i=\hat{\epsilon}_j}{\sum_{i\in\mathcal{S}\setminus\mathcal{E}}}\left[ \nu_{ij}+\hat{\epsilon}_j\sigma_i\right]n_i
+\underset{\hat{\epsilon}_i<\hat{\epsilon}_j}{\sum_{i\in\mathcal{S}\setminus\mathcal{E}}}\left[ \nu_{ij}+\hat{\epsilon}_j\sigma_i\right]n_i
-\hat{\epsilon}_j n_\mathrm{g}\ .
\end{equation}
After defining
\begin{equation}
\tilde{n}_{j,\mathrm{min}}=\underset{\hat{\epsilon}_i<\hat{\epsilon}_j}{\sum_{i\in\mathcal{S}\setminus\mathcal{E}}}\nu_{ij}n_i
\end{equation}
and some straight-forward algebraic manipulations, one obtains
\begin{equation}
F_j(n_j)=n_j
+\underset{\hat{\epsilon}_i=\hat{\epsilon}_j}{\sum_{i\in\mathcal{S}\setminus\mathcal{E}}}\nu_{ij}n_i
+\tilde{n}_{j,\mathrm{min}}
-\hat{\epsilon}_j\left( n_\mathrm{g}-\sum_{i\in\mathcal{S}\setminus\mathcal{E}}\sigma_i n_i\right) \ .
\end{equation}
The correction term $\tilde{n}_{j,\mathrm{min}}$ is the same as used by \citet{Sto18} (there denominated by $n_{j,\mathrm{min}}$).
Eliminating $n_\mathrm{g}$ with help of equation~(\ref{eq:ntot}) and by taking equation~(\ref{eq:nuclei}) into account, it follows
\begin{equation}
F_j=n_j+\underset{\hat{\epsilon}_i=\hat{\epsilon}_j}{\sum_{i\in\mathcal{S}\setminus\mathcal{E}}}\nu_{ij}n_i+\tilde{n}_{j,\mathrm{min}}-\hat{\epsilon}_j n_{\left\langle\mathrm{g}\right\rangle}\ .
\end{equation}
Because
\begin{equation}
n_{\left\langle\mathrm{g}\right\rangle}=\sum_{j\in\mathcal{E}}n_{\left\langle j\right\rangle}=\sum_{j\in\mathcal{E}}\epsilon_jn_{\left\langle\mathrm{r}\right\rangle}\ ,
\end{equation}
the objective function can also be expressed by
\begin{equation}
F_j=n_j+\underset{\hat{\epsilon}_i=\hat{\epsilon}_j}{\sum_{i\in\mathcal{S}\setminus\mathcal{E}}}\nu_{ij}n_i+\tilde{n}_{j,\mathrm{min}}-\epsilon_j n_{\left\langle\mathrm{r}\right\rangle}
\end{equation}
which is the objective function used in \texttt{FastChem 1}.
However, in contrast to \texttt{FastChem 1}, here, depending on the initial values of $n_{j,\mathrm{min}}^{(0)}$ and $n_0^{(0)}$, $n_{\left\langle\mathrm{r}\right\rangle}$ is not necessarily constant over the iteration procedure.
That is why in \texttt{FastChem 2} no additional iteration procedure is required to determine the fitting $n_{\left\langle\mathrm{r}\right\rangle}$.

\bsp	
\label{lastpage}
\end{document}